\documentclass[pre,twocolumn,showpacs,showkeys,preprintnumbers,amsmath,amssymb]{revtex4-1}
\usepackage{graphicx}
\usepackage{bm}
\begin{document}

\title{Dynamic hysteretic features of the Ising-typed thin films}

\author{Bahad\i r~Ozan~Akta\c{s}}
\affiliation{Dokuz Eyl\"{u}l University, Graduate School of Natural and Applied Sciences, TR-35160 Izmir, Turkey}
\author{{\" U}mit Ak\i nc\i}
\author{Hamza Polat}
\email{hamza.polat@deu.edu.tr}
\affiliation{Department of Physics, Dokuz Eyl\"{u}l University, TR-35160 Izmir, Turkey}

\date{\today}

\begin{abstract}
In order to elucidate the nature of hysteresis characteristics in a magnetic Ising-typed thin film with certain thickness, such as types of the frequency dispersion curves, the decay of hysteresis loop area and the corresponding coercive field and remanent magnetization values, etc., we investigate the hysteretic response of each layers within the effective-field theory. Throughout the analysis, the best appropriate parameter values are chosen since they would allow us to observe the reversed magnetic hysteresis after a certain value of external field frequency. This eccentrical phenomena has prompted us to associate itself to the domain nucleation and growth mechanism in the dynamic process. Exotic shapes of the response for different layer indices in two different regimes of modified surface exchange are particularly emphasized.
\end{abstract}


\maketitle

\section{Introduction}\label{intro}
Hysteresis is the signature of how a cooperative many-body system parameters respond dynamically to an alternating external magnetic field sweep. For kinetic Ising-typed systems, the most familiar one is in the form of Lissajous curve and essentially originated from the variation of ordinary (time-dependent) magnetization with external field. This ubiquitous phenomenon has been subjected of intensive interest due to a broad range of applications. Hysteretic behavior is, however, a complicated process of a dynamic and nonlinear nature that eludes serious treatment, both experimentally and theoretically since the area enclosed by the hysteresis loop (HL) is directly proportional to the energy loss in a magnetization-demagnetization cycle. In order to understand and clarify the behavior of the system under consideration, the remanent magnetization (in other words residual magnetization which is the magnetization left behind in the system after the external magnetic field is removed) and coercive field (which means that the intensity of the external magnetic field needed to change the sign of the magnetization) has been calculated at several times in literature by benefiting from the hysteresis. HL areal scaling law -which enables a universality classification- has been manifested first by Steinmetz \cite{steinmetz} empirically on his pioneering work dating back to the end of the nineteenth-century.

Many efforts have been devoted to make a prediction for experimental verification of the hysteresis loop area (HLA) scaling for thin films (for a brief review of HLA scaling results see Ref. \cite{suen1}). The recent series of studies on usual characteristics of the hysteretic response in thin films have been also presented by different researchers \cite{jiang, he, suen2, laosiritaworn}. The surface critical properties at a dynamic phase transition have been presented very accurately by Park and Pleimling \cite{park}. According to their findings, the kinetic surface phase diagram in three dimensions is remarkably simple and does not exhibit a special transition point, nor a surface or extraordinary transition.

Low coercivity, which means low hysteresis loss per cycle of operation, is desired in core materials (or it is only used for a temporal magnetic recording). Contrary to this, high coercivity is used in applications where the magnetic recording media is frequently used or the long-lifetime recording is needed. This strategy is the main trend-setter in magnetic thin film design \cite{osaka}. Controlling thickness of film to obtain the magnetic hysteresis at a right shape to suit desired technological purpose is the key issue in disguise \cite{suen2}. Therefore, how the spontaneous spin formation mechanism involves in these reduced structures becomes the focus of frequent investigation issues. Due to the underlying complexity of reduced dimension, any systematic investigation regarding the hysteresis behavior and its influences on each layer are not exist. So, there are some important issues which remain in suspense; frequency dispersion of HLA with corresponding remanence and coercivity for different layer indices, classification of dispersion curves, topological evolution of the hysteretic response of each layer with varying the modified surface exchange, etc.

Consequently, the aim of the study is to give a more complete picture of the overall hysteretic behavior of ferromagnetic thin films for different values of modified exchange interaction at a fixed temperature (the temperature has been selected a certain criterion). For this purpose, we first investigate how the hysteretic properties of a film depend on the dynamic system parameters by means of effective-field theory (EFT) including their relevant dispersion curves. After that, we present a sort of well-selected hysteresis collection (they have been selected precisely from the frequency dispersion of the HLA) to observe how the response depends on the layer index. Finally we discuss the magnetization reversal mechanism within reasonable bounds.

\section{Methodology}\label{method}
From the theoretical point of view, the most widely used model to study the magnetic properties of surfaces is Ising model due to the crossover of dimensionality and strong uniaxial anisotropy. So the system can be modeled by a layered Ising-typed structure which consists of interacting parallel layers. Each layer is defined as a regular lattice with coordination number $z=4$. The Hamiltonian is given by,
\begin{equation}\label{eq1}
\mathcal{H}=-\sum_{\langle ij\rangle}J_{ij}s_{i}s_{j}-h(t)\sum_{i} s_{i}
\end{equation}
where $s_{i}$ is the spin operator on lattice site $i$ and any spin variable can take the values $s_{i}=\pm 1$. $\langle \ldots\rangle$ subscript bracket symbolizes the nearest neighboring in the first summation. The second summation is over all the lattice sites. The exchange interaction $J_{ij}$ between the spins on the sites $i$ and $j$ take the values according to the positions of the nearest neighbor spins. Two surfaces of the film have intralayer coupling $J_{1}$. The interlayer coupling between the surface and its adjacent layer (i.e. layers $1,$ $2$ and $L,$ $L-1$) is denoted by $J_{2}$. For the rest of the layers, the interlayer and the intralayer couplings are assumed as $J_{3}$. The system has three exchange interactions where $J_{1},$ $J_{2},$ $J_{3}>0$ favors a ferromagnetic alignment of the adjacent sites as shown in Fig. (1) of our prior work (see in Ref. \cite{aktas1}). The Zeeman term describes interaction of the spins with the field of the sinusoidal form
\begin{equation}\label{eq2}
h(t)=h_{0}\cos(\omega t),
\end{equation}
where $t$ is the time and $h_{0}$ is the amplitude of the oscillatory magnetic field with an angular frequency $\omega$.

Our system is in contact with an isothermal heat bath at given temperature $T$. So, the dynamical evolution of the system may be given by non-equilibrium Glauber dynamics \cite{glauber} based on a master equation. Number of $L$ different magnetization time series for the system can be given by usual dynamical EFT equations which are obtained by differential operator technique \cite{honmura, kaneyoshi1}. In order to handle the multi-spin correlations, a decoupling approximation (DA) \cite{tucker} can be used as
\begin{equation}\label{eq3}
\langle s_{i}^{(k)}\ldots s_{j}^{(k)}\ldots s_{l}^{(k)}\rangle=\langle s_{1}^{(k)}\rangle\ldots \langle s_{j}^{(k)}\rangle \ldots \langle s_{l}^{(k)}\rangle,
\end{equation}
which is essentially identical to the Zernike approximation \cite{zernike} in the bulk problem, and it has been successfully applied to a great number of magnetic systems including the surface problems \cite{balcerzak, kaneyoshi2, kaneyoshi3}. Thus, the dynamical equations of motion for each layer are in the form
\begin{widetext}
\begin{eqnarray}\label{eq4}
\tau \frac{dm_{1}}{dt}&=&-m_{1}+[A_{1}+m_{1}B_{1}]^{z}[A_{2}+m_{2}B_{2}],\nonumber\\
\tau \frac{dm_{2}}{dt}&=&-m_{2}+[A_{2}+m_{1}B_{2}][A_{3}+m_{2}B_{3}]^{z}[A_{3}+m_{3}B_{3}],\nonumber\\
\vdots\quad&\vdots&\quad\quad\quad\quad\quad\quad\quad\quad\vdots\nonumber\\
\tau \frac{dm_{k}}{dt}&=&-m_{k}+[A_{3}+m_{k-1}B_{3}][A_{3}+m_{k}B_{3}]^{z}[A_{3}+m_{k+1}B_{3}],\nonumber\\
\vdots\quad&\vdots&\quad\quad\quad\quad\quad\quad\quad\quad\vdots\nonumber\\
\tau \frac{dm_{L-1}}{dt}&=&-m_{L-1}+[A_{3}+m_{L-2}B_{3}][A_{3}+m_{L-1}B_{3}]^{z}[A_{2}+m_{L}B_{2}],\nonumber\\
\tau \frac{dm_{L}}{dt}&=&-m_{L}+[A_{2}+m_{L-1}B_{2}][A_{1}+m_{L}B_{1}]^{z}.
\end{eqnarray}
\end{widetext}
Here, the magnetization time series of $k$th layer $m_{k}$ is defined as
\begin{equation}\label{eq5}
m_{k}=\langle s_{i}^{(k)}\rangle,\quad k=1,\ldots,L,
\end{equation}
with the coefficients
\begin{eqnarray}\label{eq8}
A_{n}&=&\cosh(J_{n}\nabla)f(x)|_{x=0},\nonumber\\
B_{n}&=&\sinh(J_{n}\nabla)f(x)|_{x=0}.
\end{eqnarray}
$n=1,2,3$ and $\nabla=\partial/\partial x$ is one dimensional differential operator and the function $f(x)$ is given by
\begin{equation}\label{eq9}
f(x)=\tanh[\beta(x+h(t))].
\end{equation}
Here, $\beta=1/k_{B}T$ and $k_{B}$ represents the Boltzmann constant. The effect of the differential operator on an arbitrary function $f(x)$,
\begin{equation}\label{eq10}
\exp(a\nabla)f(x)|_{x=0}=f(x+a)|_{x=0},
\end{equation}
with any real constant $a$.

By using the binomial expansion and the hyperbolic trigonometric functions in terms of the exponential functions we get the most compact form of Eq. (\ref{eq4})
\begin{widetext}
\begin{eqnarray}\label{eq11}
\dot{m}_{1}&=&\frac{1}{\tau}\left(-m_{1}+\sum_{\gamma=0}^{z}\sum_{\eta=0}^{1}\Lambda_{1}(\gamma,\eta)m_{1}^{\gamma}m_{2}^{\eta}\right),\nonumber\\
\dot{m}_{2}&=&\frac{1}{\tau}\left(-m_{2}+\sum_{\gamma=0}^{z}\sum_{\eta=0}^{1}\sum_{\nu=0}^{1}\Lambda_{2}(\gamma,\eta,\nu)m_{2}^{\gamma}m_{1}^{\eta}m_{3}^{\nu}\right),\nonumber\\
\vdots\quad&\vdots&\quad\quad\quad\quad\quad\quad\quad\quad\quad\quad\vdots\nonumber\\
\dot{m}_{k}&=&\frac{1}{\tau}\left(-m_{k}+\sum_{\gamma=0}^{z}\sum_{\eta=0}^{1}\sum_{\nu=0}^{1}\Lambda_{3}(\gamma,\eta,\nu)m_{k}^{\gamma}m_{k-1}^{\eta}m_{k+1}^{\nu}\right),\nonumber\\
\vdots\quad&\vdots&\quad\quad\quad\quad\quad\quad\quad\quad\quad\quad\vdots\nonumber\\
\dot{m}_{L-1}&=&\frac{1}{\tau}\left(-m_{L-1}+\sum_{\gamma=0}^{z}\sum_{\eta=0}^{1}\sum_{\nu=0}^{1}\Lambda_{2}(\gamma,\eta,\nu)m_{L}^{\gamma}m_{L-1}^{\eta}m_{L+1}^{\nu}\right),\nonumber\\
\dot{m}_{L}&=&\frac{1}{\tau}\left(-m_{L}+\sum_{\gamma=0}^{z}\sum_{\eta=0}^{1}\Lambda_{1}(\gamma,\eta)m_{L}^{\gamma}m_{L-1}^{\eta}\right),
\end{eqnarray}
\end{widetext}
where
\begin{eqnarray}\label{eq12}
\Lambda_{1}(\gamma,\eta)&=&{{z}\choose{\gamma}}{{1}\choose{\eta}}A_{1}^{z-\gamma}A_{2}^{1-\eta}B_{1}^{\gamma}B_{2}^{\eta},\nonumber\\
\Lambda_{2}(\gamma,\eta,\nu)&=&{{z}\choose{\gamma}}{{1}\choose{\eta}}{{1}\choose{\nu}}A_{2}^{1-\eta}A_{3}^{z+1-\gamma-\nu}B_{2}^{\eta}B_{3}^{\gamma+\nu},\nonumber\\
\Lambda_{3}(\gamma,\eta,\nu)&=&{{z}\choose{\gamma}}{{1}\choose{\eta}}{{1}\choose{\nu}}A_{3}^{z+2-\gamma-\eta-\nu}B_{3}^{\gamma+\eta+\nu}.
\end{eqnarray}
Here, $1/\tau$ is transition per unit time in a Glauber type stochastic process and throughout our calculations we set it as $\tau=1$ for simplicity. It is clear that solving the set of self-consistent nonlinear ordinary differential equations given in Eq. (\ref{eq11}) is essentially initial value problem. In order to get the $m_{k}(t)$ time-series for each, we use standard fourth order Runge-Kutta method (RK4).

The system has three dependent Hamiltonian variables, namely frequency and amplitude of external magnetic field respectively $\omega$, $h_{0}$, and the thickness $L$. For fixed values of these parameters together with temperature and interaction constants $J_{1},$ $J_{2},$ $J_{3}$, RK4 solutions will give convergency behavior after some iterations, i.e., the solutions have property $m(t)=m(t+2\pi/\omega)$ for arbitrary initial value for the magnetization ($m(t=0)$). Each iteration, i.e., the calculation of magnetization at $t+1$ time step from previous magnetization for $t$, is now performed for these purpose whereby the RK4 iterative equation is being utilized to determine the magnetization for every $i$. In order to keep the iteration procedure stable in our simulations, we have chosen $10^{4}$ points for each RK4 step. Thus, after obtaining the convergent region and some transient steps (which depends on Hamiltonian parameters and the temperature) the dynamical order parameter (DOP) for each layers can be calculated from
\begin{equation}\label{eq13}
Q_{k}=\frac{\omega}{2\pi}\oint m_{k}(t)dt
\end{equation}
where $m_{k}$ is a stable and periodic function anymore.

As cut-off condition for numerical self-consistency, we defined a tolerance
\begin{equation}\label{eq14}
\left |Q_{k}|_{t-2\pi/\omega}^{t}-Q_{k}|_{t}^{t+2\pi/\omega}\right |<10^{-5},
\end{equation}
meaning that the maximum error as difference between the each consecutive iteration should be lower than $10^{-5}$ for all step. DOP accurately can be calculated from this stationary solutions of $m_{k}(t)$ over a cycle. On the other hand, the energy loss due to the hysteresis, in other words, HLA is defined as
\begin{equation}\label{eq15}
A^{(k)}=-\oint m_{k}(t)dh(t)=h_{0}\omega\oint m_{k}(t)\sin(\omega t),
\end{equation}
and the modified surface exchange interaction has been defined to determine the different characteristic behavior of the system in certain range as
\begin{equation}\label{eq16}
J_{1}=J_{3}(1+\Delta_{s}).
\end{equation}

There are three possible order of the system: F, P and the coexistence phase (F+P). In P phase, $m(t)$ in convergent region for any layer is satisfied by this condition
\begin{equation}\label{eq17}
m(t)=-m(t+\pi/\omega)
\end{equation}
which is called the symmetric solution. On the other hand, in F phase, the solution does not satisfy Eq. (\ref{eq17}) and this solution is called non-symmetric solution which oscillates around a non-zero magnetization value. The solution in F phase does not follow the external magnetic field, i.e., the value of $Q$ is different from zero. In these two cases, the observed behavior of magnetization is independent from the choice of initial value of magnetization $m(0)$ whereas the last phase has magnetization solutions symmetric or non-symmetric depending on the choice of the initial value of magnetization corresponding to the coexistence region where F and P phases overlap. The main goal of our detailed investigation is adjunctly manifesting the frequency dispersion of HLA and corresponding coercivity $(H^{(k)}_{c})$ with remanence $(M^{(k)}_{r})$ of layers. Note that, more details about formulation and the reasons for selection procedure about the related values of system parameters at which the calculations carried out can be seen in Ref. \cite{aktas1}.

\section{Results and Discussion}\label{result}
\begin{figure}
\centering
\includegraphics[width=7.8cm]{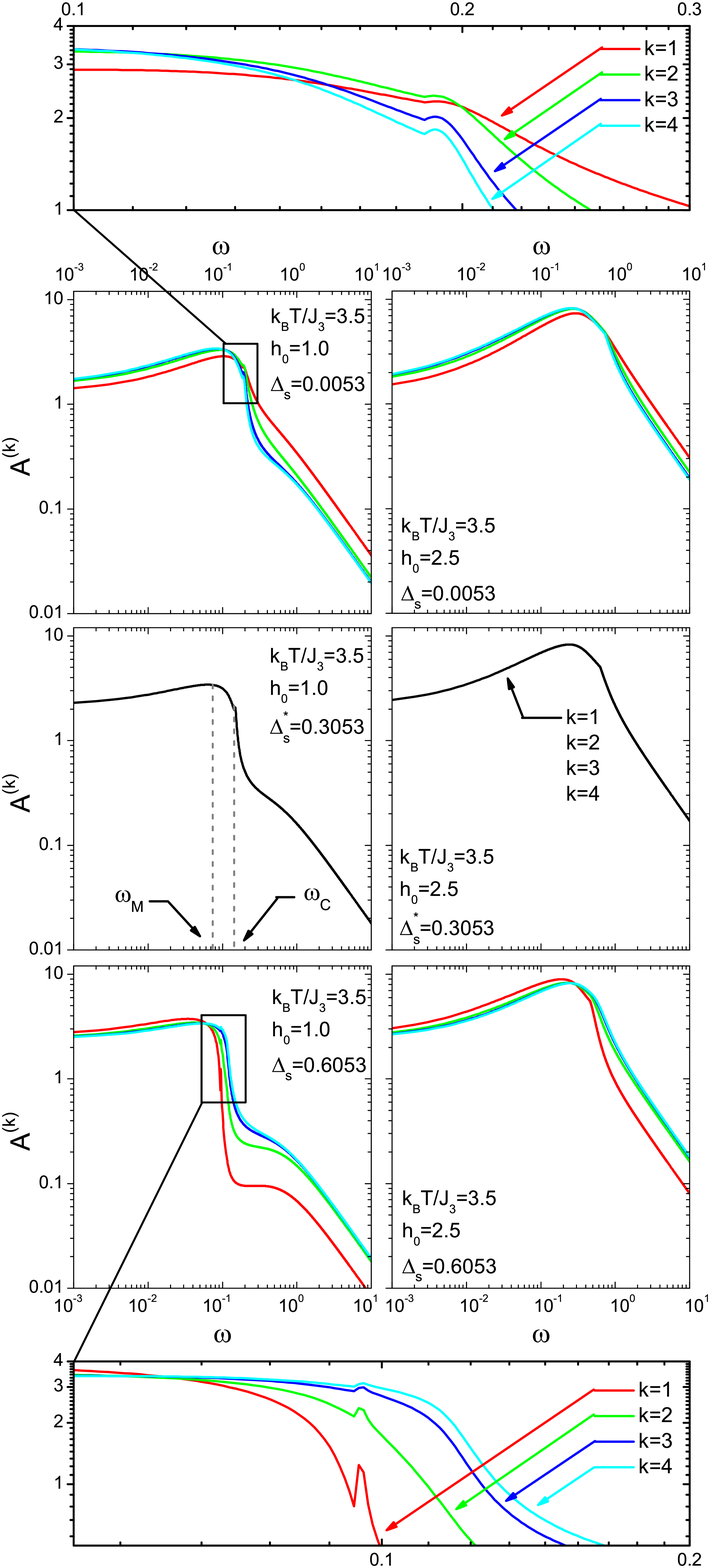}
\caption{(Color online) The frequency dispersion of HLA for each layer of $L=7$ at a fixed temperature $k_{B}T/J_{3}=3.5$. The curves were plotted for the selected amplitude representatives as $h_{0}=1.0$ and $2.5$ in each group of triple-column. The sequence of dispersions is reversed in two different regimes of modified exchange interaction.}\label{fig1}
\end{figure}

\begin{figure}
\centering
\includegraphics[width=7.82cm]{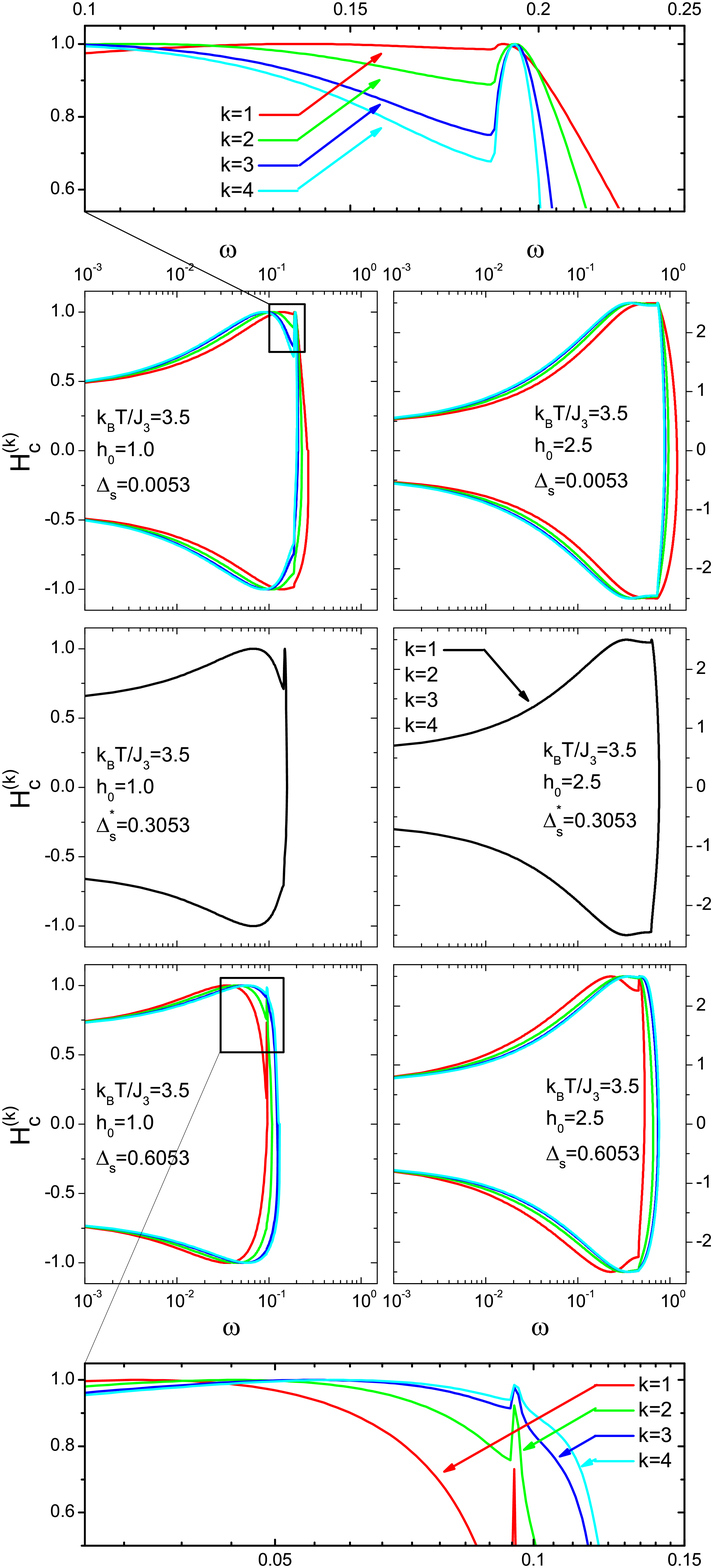}
\caption{(Color online) The frequency dispersion of coercive field for the same selected system parameters of the film depicted in Fig. (\ref{fig1}). The sequence of dispersions is reversed in two different regimes of modified exchange interaction.}\label{fig2}
\end{figure}

\begin{figure}
\centering
\includegraphics[width=8.3cm]{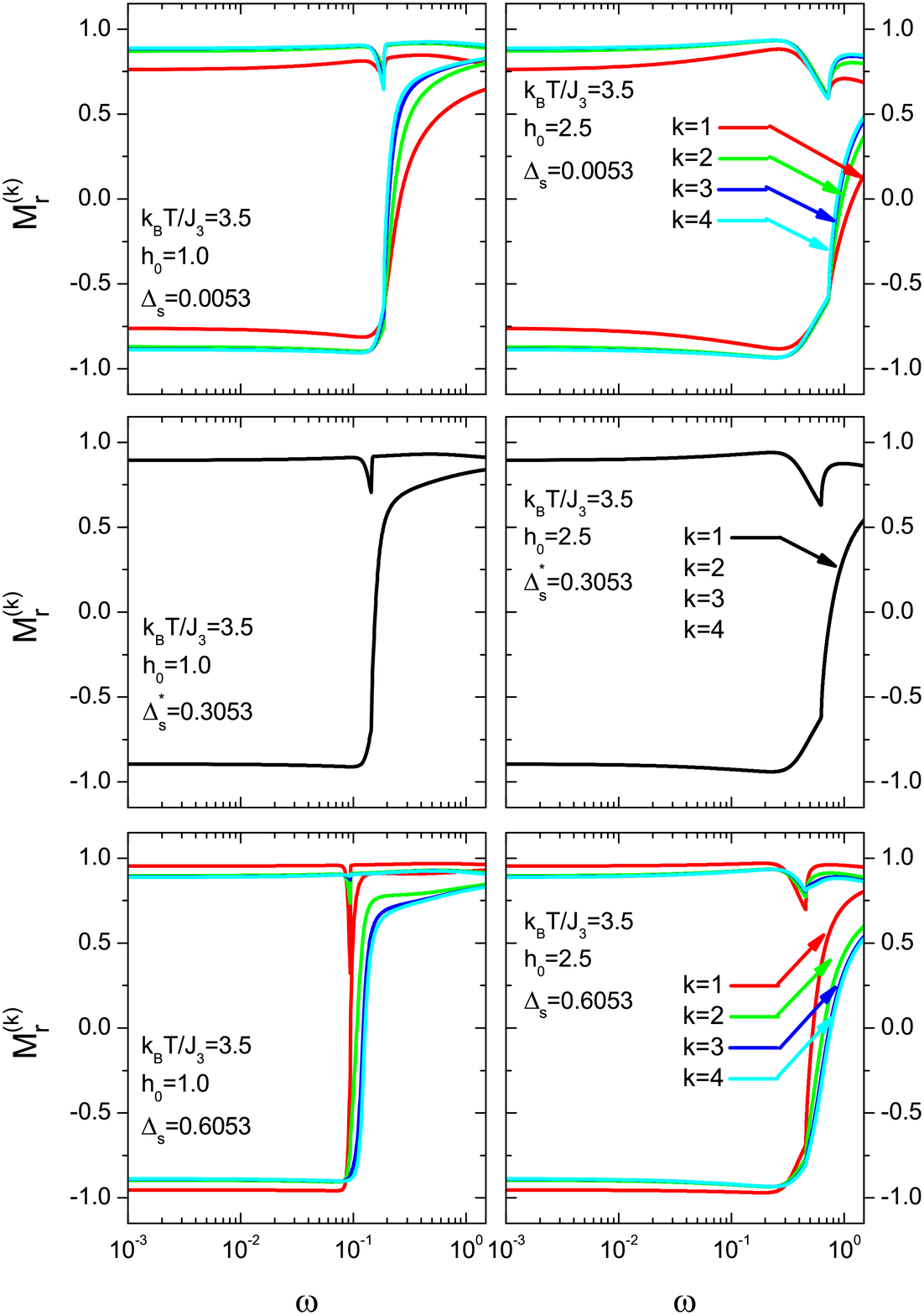}
\caption{(Color online) The frequency dispersion of remanent magnetization for the same selected system parameters of the film depicted in Fig. (\ref{fig1}). The sequence of dispersions is reversed in two different regimes of modified exchange interaction.}\label{fig3}
\end{figure}

With the guidance of the work given in Ref. \cite{aktas1}, the frequency dispersion of HLA with corresponding coercivity and remanent magnetization for a pure crystalline ferromagnetic thin film with $L=7$ layers are presented in Figs. (\ref{fig1}), (\ref{fig2}) and (\ref{fig3}). We see that the frequency dependency in each characteristic has a crossover at a critical value of modified exchange interaction $\Delta_{s}^{*}$. There exists a hierarchical change of the HLA sequence for each layer in two different regimes of $\Delta_s$ as shown in Fig. (\ref{fig1}). For a fixed $h_{0}$, the hysteresis loop of surface has the lowest area in $\Delta_{s}<\Delta_{s}^{*}$ regime till the symmetry loss. In other regime ($\Delta_{s}>\Delta_{s}^{*}$) the hysteresis loop of innermost layer has the lowest area. The layer indices are of no importance in frequency dependency at the critical value of modified exchange interaction. In other words, all the layers have the same hysteretic behavior at $\Delta_{s}^{*}$. Physical mechanism can be briefly explained as follows: When proceeding from the surface to the center of the film, hysteresis loop becomes larger at lower $\omega$ but it slightly gets smaller at higher $\omega$ values for the fixed value of $h_{0}$ in $\Delta_{s}<\Delta_{s}^{*}$ regime. There are relatively more neighboring per magnetic-sites which causes locally larger magnetic interaction in inner layers. So, it becomes more difficult to follow the external field for any spin. Hysteresis is more likely to be asymmetric in this regime. Surface spins are embedded in an environment of lower symmetry than that of the inner atoms. The exchange constant between atoms in the surface region may differs from the bulk one. The lower $\Delta_{s}$ regime corresponds to surface type of magnetic ordering with this aspect. The opposite of the above scenarios can be considered also. From a different point of view, if the $\Delta_{s}$ increases (i.e. we are in $\Delta_{s}>\Delta_{s}^{*}$ regime now), hysteresis loop becomes smaller at low $\omega$ but slightly gets larger at high $\omega$ when proceeding from the surface to the center of the film. This is since for inner layers there are more neighboring but far fewer exchange constant per magnetic-sites which causes relatively smaller magnetic interaction than that of the surface one anymore. Eventually, the free surface can not breaks the translational symmetry since the hysteretic and/or magnetic properties of the free surfaces exactly overlap with the bulk one at the critical value of surface exchange $\Delta_{s}^{*}$. We can say more generally that the deficiency of the interaction strength per surface spin can be compensated by increasing the modified exchange interaction strength. The coercivity and remanent magnetization presented in Figs. (\ref{fig2}) and (\ref{fig3}) are also consistent with the aforementioned arguments about the mechanism. They profiles across the film differ qualitatively in two regimes ($\Delta_{s}< \Delta_{s}^{*}$ and $\Delta_{s}> \Delta_{s}^{*}$).


$A_{k}$ increases at low $\omega$ values first and then it reaches the maximum value in agreement with the well-known behavior in literature where frequency at the peak $\omega_{M}$ corresponds to the phase-lag. After $\omega_{M}$, the hysteresis is in still purely symmetric region with reversed shape for a while. In $\Delta_{s}<\Delta_{s}^{*}$ regime, $\omega_{M}$ shifts to lower values with increasing layer index $k$ (it means we are penetrating from the surface to the center) due to the locally stronger magnetic interaction. $A_{k}$ is also increases (especially at low $\omega$) with increasing layer index $k$ for fixed $\omega$ in the same regime of $\Delta_{s}$. The opposite of all these arguments can be also thought for $\Delta_{s}>\Delta_{s}^{*}$ regime. One can easily observe the intersection of layers via changing $\Delta_{s}$ parameter at $\omega_{M}$.

One can easily follow from Figs. (\ref{fig2}) and (\ref{fig3}) respectively, the coercivity $H_{c}^{(k)}$ is relatively large for larger field amplitude $h_{0}$. The effect of larger field amplitude $h_{0}$ on remanence $M_{r}^{(k)}$ can be seen also in asymmetric region. $A_{k}$ increases with increasing $h_{0}$ and this increment is independent of the other system parameters. This is since the magnetic energy due to the energy supplied by the external field in a cycle automatically becomes higher for higher amplitude values. This higher energy provides more magnetic force which increases the ability of the spins to follow the external field sweep. Consequently, corresponding frequency for the maximal peak ($\omega_{M}$) in symmetric region and value of $\Delta_{s}$ shifts to higher frequencies. It is clear that the phase-lag between magnetization time-series and external field signal is smaller for the higher amplitude.

\begin{figure*}
\centering
\includegraphics[width=17.5cm]{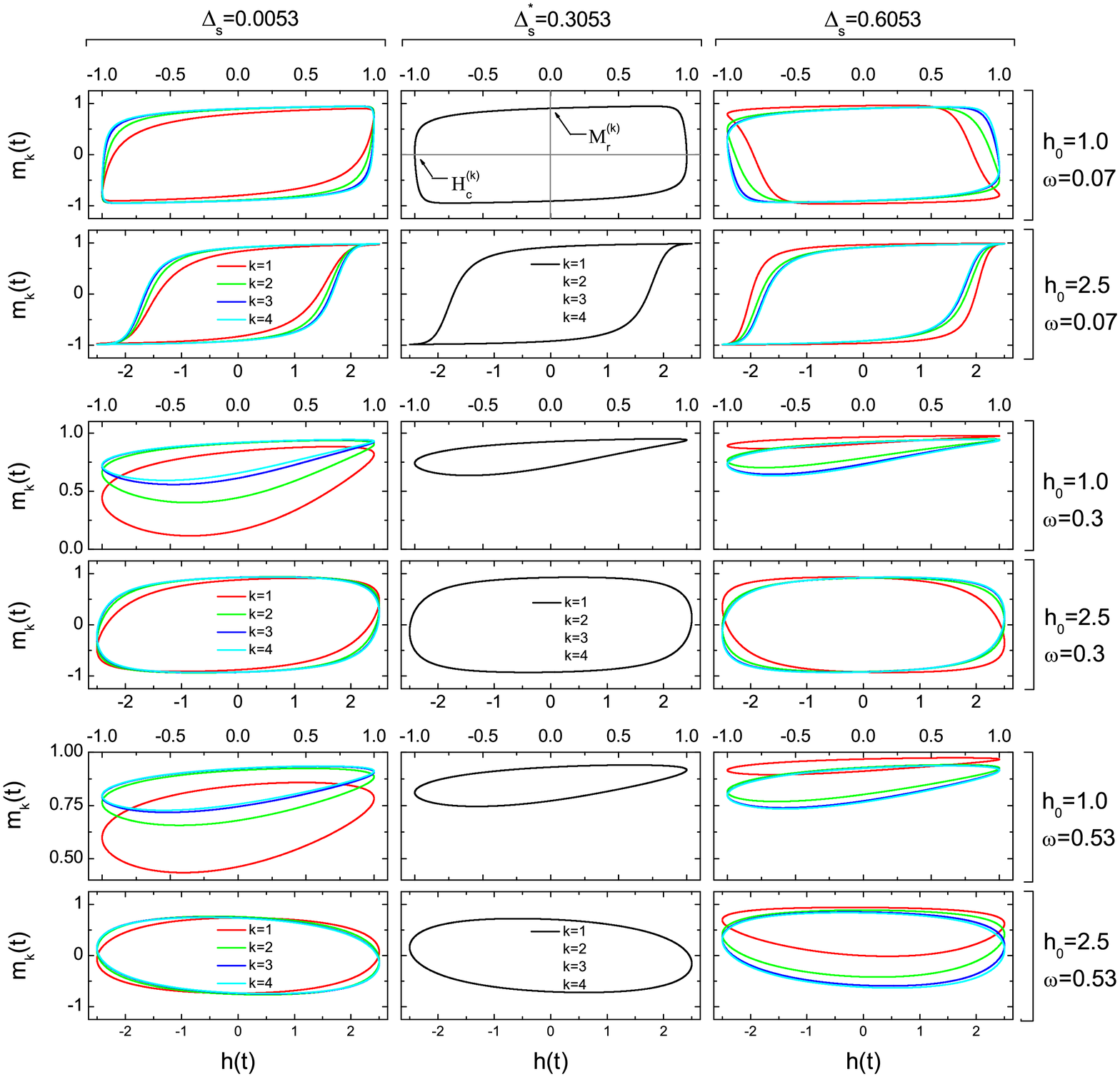}
\caption{(Color online) Well-selected hysteresis collection of each layer for the film with thickness $L=7$. Modified exchange interaction $\Delta_{s}$ ranges from 0.0053 (first column) to 0.3053 (second column) and 0.6053 (third column) with varying frequency as $\omega=0.07$, $0.3$ and $0.53$ for related amplitude representatives at $k_{B}T/J_{3}=3.5$. The reader can recognize the hierarchical evolution in grayscale version of print by following the Supplemental Material given in in Ref. \cite{supp}. Here, we note that the surface loop ($k=1$) is more saturated in any case for $\Delta_{s}<\Delta_{s}^{*}$ value and the sequence of hysteresis is reversed in two different regimes of modified exchange interaction.}\label{fig4}
\end{figure*}

The hysteresis loop collection of each layer for the selected system parameters together with three appropriate values of frequency are presented in Fig. (\ref{fig4}). In the low frequency limit, the magnetization can be more saturated and this shows itself as dominant magnetization processes in the different regions of the hysteresis (i.e. reversible boundary displacements, irreversible boundary displacements and magnetization rotation). The figure shows that the hysteretic response of each layer clearly has the same characteristics at the critical value of the modified exchange $\Delta_{s}^{*}$ independently from the other parameters. Hierarchically changing the order of the hysteresis loop for various values of amplitudes and/or frequency also can be seen in Fig. (\ref{fig4}) by following the group of three columns including the related regimes of $\Delta_{s}$ ($\Delta_{s}<\Delta_{s}^{*}$, $\Delta_{s}\equiv\Delta_{s}^{*}$ and $\Delta_{s}>\Delta_{s}^{*}$). One can observe the symmetry process on this order changing (corresponds to the following ability of the spin to external field sweep) by comparing the representative amplitudes between each other. For $Delta_{s}^{*}$ value and $h_{0}=2.5$ representative, the hysteresis loop has a saturated s-shape and tends to enhance its size with increasing $\omega$ at low frequency region e.g. $\omega<0.3$. However, on further increasing $\omega$, the loop gets its maximum area and later on reduces to an oval shape with its major axis parallel to the field axis. This is the result of phase-lag between the magnetization time-series and the field signal. At very low $\omega$, the field period is large and the magnetic spins have sufficient time to follow the field signal so the phase-lag between the magnetization time-series and field signal is very small. The system is strong enough to follow the external perturbation at those limitations, hence the hysteresis loop looks like a slim s-shape. However, on increasing $\omega$, the field sweeps faster. But, the relaxation time of the system is fixed, so spins still have less time to follow the field. Phase-lag gets larger and so does the HLA. While the frequency value is approaching to correspondence phase-lag, the hysteresis gets its maximum size which corresponds to the frequency $\omega_{M}$. After that, if $\omega$ still increases in small doses, the system can not follow the field and the spins do not align in the same direction with external oscillatory field in each moment of a full cycle. Finally, the hysteresis dramatically loses its symmetry at a certain $\omega_{C}$ value. For $h_{0}=1.0$, a rare hysteresis shape and symmetry loss with increasing frequency can be seen by following the aforementioned topological evolution at the fixed $\Delta_{s}>\Delta_{s}^{*}$. For instance, as seen at $\omega=0.07$, an exotic left-handed s-shape beyond the oval form temporally takes place between corresponding maximal area frequency $\omega_{M}$ and the frequency value of symmetry loss $\omega_{C}$ where $\omega_{C}>\omega_{M}$. The general trend mainly can be also explained as follow: For low frequencies, the hysteresis loop is in the form of saturated s-shape and tends to enhance its size with increasing $\omega$ at low frequency region. At $\omega_{M}$ value, the loop gets its maximum area. With a further increase in frequency, hysteresis enters into the process of smooth directional-veering and it changes its shape to left-handed s-shape until the symmetry loss. Finally, the symmetry loss occurs at a characteristic frequency value $\omega_{C}$, so the hysteresis has asymmetric shape anymore.

The process stated between at the beginning of directional-veering and symmetry loss (in between $\omega_{M}$ and $\omega_{C}$) deserves a scrutiny in terms of magnetic domain nucleation. As seen in Figs. (\ref{fig1}), (\ref{fig2}) and (\ref{fig3}), symmetry loss shows itself as a gibbous-like eminentia in $\omega_{C}$. This phenomenon is also zoomed in Figs. (\ref{fig1}) and (\ref{fig2}) for better visibility. Particularly, as seen in Fig. (\ref{fig3}), both positive and negative remanence appears at low frequencies, but only positive values survive at high frequencies. This is also another indication of symmetry loss. The related eminentia shows itself more clearly on right arm of the mutual coercivity and it is more and more distinct in surfaces for low amplitudes especially in $\Delta_{s}>\Delta_{s}^{*}$ regime. We note that the encountered characteristics on this problem are the member of Type II.b family of frequency dispersion presented by Akta\c{s}~et al. \cite{aktas2, supp}. As we mentioned above, low amplitude relatively provides smaller magnetic energy to the system and this weak energy could not provide enough magnetic force in causing the spins to follow the field. This means that the surfaces fall all over itself to adopt to the asymmetric phase in the $\Delta_{s}>\Delta_{s}^{*}$ regime. So, the surfaces have extremely left-handed saturated s-shape hysteretic response within that period. It is also an special interest to consider the topological evolution of the hysteresis with varying $\omega$ to investigate how it relates with spin formation (magnetic domains) growth and/or nucleation mechanism. Two branches of the hysteresis are almost lie on two perpendicular directions at $\omega_{M}$. HLA has its maximum area which is related to these two axis and their orientation. Therefore, the magnetic moments remain a constant value in the half-time period of external field at this frequency $\omega_{M}$. At the fixed values of $\omega<\omega_{M}$ and $\omega>\omega_{M}$ respectively, corresponding hysteresis curves are topologically contra-oriented and completely different growth mechanisms are onset. The importance of this left-handed shape on the domain mechanism is that the domains currently continue to nucleate in a direction with a delay, while the magnetic field is increasing in the opposite direction in a half period of full cycle. This phenomenon should be considered especially on the fabrication process of thin films for magnetic recording purposes. Magnetization time-series versus the external magnetic field, increasing the field frequency at first, obstructs the saturation of the ordinary magnetization due to the decreasing energy coming from the oscillating magnetic field in the half-time period which facilitates the late stage domain growth by tending to align the moments in its direction (i.e. the magnetization is no longer able to follow the oscillatory field). This enables a frequency increasing route to continuous dynamical phase transition due to the incomplete reversal of magnetic moments. The aforementioned directional-veering mechanism begins to lose its significance in the $\Delta_{s}<\Delta_{s}^{*}$ regime. When the $\Delta_{s}$ is lowered more and more, the dipole-dipole interaction-induced energy contribution becomes smaller. The strength of the energy contribution which comes from the dipole-dipole interaction corresponds to remain in dynamically symmetric phase of the system's own volition. Hence, the surfaces can stay in the symmetric phase in the smaller frequency until the frequency of the external field becomes greater than the intrinsic relaxation time of the system.

\section{Conclusion}\label{conclude}
The effect of modified exchange interaction on dynamical Ising-typed thin film driven by an external oscillatory magnetic field is discussed by means of EFT based on standard DA. The time evolution of the system is presented by utilizing a Glauber type stochastic process. The investigation is focused on frequency dispersion of HLA, coercivity and the remanence of each layer.

In order to find out the physical relation between domain growth and/or nucleation mechanism and symmetry loss, the topological evolution of the hysteresis with varying frequency is considered. Hence, some unpretentious improvements are developed in accordance with the explanations given before in literature \cite{jiang, he, suen2, laosiritaworn}.

The existence of a special transition point presented in Ref \cite{aktas1} has been already confirmed by Tauscher and Pleimling via both Glauber and Metropolis dynamics \cite{tauscher}. Moreover, one can find to be of another interest is the existence of a critical value of the surface exchange parameter at which all the layers seem to oscillate in phase. Based on the Monte Carlo results \cite{tauscher}, we can say that the existence a crossover in all hysteretic characteristics are not from the limitation of the method. This would hopefully provide some theoretical insight into the role of the surface layer and possibly some guidance to experimental studies on such systems.

Especially the short-time reversed shape of hysteresis in the frequency-induced evolution which can only be recognized with such detailed examination as obtained from this work is another issue worth considering. We hope that, this study would shed light on the further investigations of the dynamic nature of critical phenomena in pure crystalline ferromagnetic thin films and would be beneficial from both theoretical and experimental points of view.

\section*{Acknowledgements}
The numerical calculations in this paper were performed at T\"{U}B\.{I}TAK ULAKB\.{I}M (Turkish agency), High Performance and Grid Computing Center (TRUBA Resources) and this study was completed at Dokuz Eyl\"{u}l University, Graduate School of Natural and Applied Sciences. One of the authors (B.O.A.) would like to thank the Turkish Educational Foundation (TEV) for full scholarship.


\begin{thebibliography}{99}
\bibitem{steinmetz} C. P. Steinmetz, Trans. Am. Inst. Electr. Eng. \textbf{9}, 3 (1892).
\bibitem{suen1} J. S. Suen and J. L. Erskine, Phys. Rev. Lett. \textbf{78}, 3567 (1997).
\bibitem{jiang} Q. Jiang, H. N. Yang, and G. C. Wang, Phys. Rev. B \textbf{52}, 14911 (1995).
\bibitem{he} Y. L. He and G. C. Wang, Phys. Rev. Lett. \textbf{70}, 2336 (1993).
\bibitem{suen2} J. S. Suen, M. H. Lee, G. Teeter and J. L. Erskine, Phys. Rev. B \textbf{59}, 4249 (1999).
\bibitem{laosiritaworn} Y. Laosiritaworn, Thin Solid Films \textbf{517}, 5189 (2009).
\bibitem{park} H. Park and M. Pleimling, Phys. Rev. Lett. \textbf{109}, 175703 (2012).
\bibitem{osaka} T. Osaka, T. Asahi, J. Kawaii, and T. Yokoshima, Electrochim. Acta \textbf{50}, 4576 (2005).
\bibitem{aktas1} B. O. Akta\c{s}, {\" U}. Ak\i nc\i, and H. Polat, Thin Solid Films \textbf{562}, 680 (2014).
\bibitem{glauber} R. J. Glauber, J. Math. Phys. \textbf{4}, 294 (1963).,
\bibitem{honmura} R. Honmura and T. Kaneyoshi, J. Phys. C: Solid State Physics \textbf{12}, 3979 (1979).
\bibitem{kaneyoshi1} T. Kaneyoshi, Acta Phys. Pol. A \textbf{83}, 703 (1993).
\bibitem{tucker} J. W. Tucker, J. Magn. Magn. Mater. \textbf{102}, 144 (1991).
\bibitem{zernike} F. Zernike, Physica A \textbf{7}, 565 (1940).
\bibitem{balcerzak} T. Balcerzak,  J. Magn. Magn. Mater. \textbf{97}, 152 (1991).
\bibitem{kaneyoshi2} T. Kaneyoshi, R. Honmura, I. Tamura, and E. F. Sarmento, Phys. Rev. B \textbf{29}, 5121 (1984).
\bibitem{kaneyoshi3} T. Kaneyoshi, I. Tamura, and E. F. Sarmento, Phys. Rev. B \textbf{28}, 6491 (1983).
\bibitem{aktas2} B. O. Akta\c{s}, {\" U}. Ak\i nc\i, and H. Polat, Physica B \textbf{407}, 4721 (2012).
\bibitem{supp} See Supplemental Material at [URL will be inserted by publisher] for Type II.b class of frequency dispersion.
\bibitem{tauscher} K. Tauscher and M. Pleimling, Phys. Rev. E \textbf{89}, 022121 (2014).
\end{thebibliography}
\end{document}